# On resonant scatterers as a factor limiting carrier mobility in graphene

Z. H. Ni[1], L. A. Ponomarenko[1], R. R. Nair[1], R. Yang[1], S. Anissimova[1], I. V. Grigorieva[1], F. Schedin[1], Z. X. Shen[2], E. H. Hill[1], K. S. Novoselov[1], A. K. Geim[1]

[1]Centre for Mesoscience & Nanotechnology, University of Manchester, Manchester M13 9PL, UK
[2]School of Physical & Mathematical Sciences, Nanyang Technological University, 637371, Singapore

*We show that graphene deposited on a substrate has a non-negligible density of atomic scale defects. This is evidenced by a previously unnoticed D peak in the Raman spectra with intensity of ~1% with respect to the G peak. We evaluated the effect of such impurities on electron transport by mimicking them with hydrogen adsorbates and measuring the induced changes in both mobility and Raman intensity. If the intervalley scatterers responsible for the D peak are monovalent, their concentration is sufficient to account for the limited mobilities currently achievable in graphene on a substrate.*

PACS numbers: 72.80.Vp, 73.22.Pr, 81.05.ue

The most important parameter that defines the range of electronic phenomena accessible in experiments on graphene is arguably its charge carrier mobility $\mu$. In graphene monolayers mechanically cleaved and deposited on a substrate, $\mu$ are usually limited to values $\approx 2$ m$^2$/Vs, which are large enough to allow ballistic transport on a submicron scale and the observation of, for example, the quantum Hall effect at room temperature (*T*) [1]. However, in search for new physics, device applications and, especially, many-body phenomena it is essential to improve the electronic quality of graphene devices further, to $\mu$ possibly as high as $10^3$ m$^2$/Vs at liquid-helium *T* [2]. To achieve this, it is important to determine what defects limit $\mu$.

The choice of available candidates is currently restricted to three: charged impurities [3,4], random strain (or ripples) [5] and resonant scatterers (RS) [6]. These are the mechanisms that result in graphene's conductivity $\sigma$ being (almost) proportional to carrier concentration *n* as observed experimentally. Many groups have argued that the dominant scatterers are charged impurities (for example, [4,7,8]). Whereas there exists a consensus that charged impurities are responsible for electron and hole puddles at the neutrality point (NP) [1,4,9] and can dominate electron transport in devices with low $\mu$ <0.1 m$^2$/Vs [8,10], some other observations suggest that Coulomb scatterers are a contributing rather than the limiting mechanism [10,11]. This issue remains controversial awaiting further clarification. A similar uncertainty surrounds the suggestion [5] that nanoscale ripples could be the limiting factor. Graphene deposited on a substrate is rippled but the observed surface topography does not comply with the strain distribution required to result in $\sigma \propto n$ [12]. Nevertheless, one can imagine that the topography induced by the SiO$_2$ roughness conceals the 'long range' strain generated by quenched flexural phonons [5]. The recent report [13] of $\mu \approx 10$ m$^2$/Vs at room *T* in suspended graphene seems to rule out the ripple mechanism. As for RS, they have become a viable candidate only recently when it was shown that hydrogen adsorbates (H) [14] and vacancies [15] led to a $\sigma(n)$ dependence indistinguishable from the typical curves for pristine graphene. RS are atomic-scale defects that generate so-called midgap states with an energy level $\varepsilon$ very close to the Dirac point [6]. Despite their atomic size, such defects result in a large scattering cross section if the Fermi energy $E_F$ is close to $\varepsilon$. When $E_F$ shifts away from $\varepsilon$ and, therefore, from the NP, the scattering rate decays relatively slow, resulting in $\sigma \propto n$ [6,16,17]. Importantly, all strongly-bound monovalent adsorbates on graphene are expected to create midgap states [16,17]. On one hand, it is easy to imagine a certain concentration of such adsorbates (e.g., OH groups). On the other hand, there has been a strong argument against RS playing any significant role: Due to their atomic scale, RS should also lead to intervalley scattering and, therefore, give rise to a D band in the Raman spectrum [18]. No such peak has been reported for pristine graphene (see, e.g., [18,19]) which seems to rule out RS as the limiting mechanism.

In this Letter, we show that, albeit unnoticeable under typical noise in Raman measurements, the D peak is universally present in graphene devices, reaching typically 1% in intensity with respect to the G peak. The D peak does not disappear after annealing at $T$ up to 400°C, which implies that the defects (further referred to as X centers) are either structural ones (e.g., vacancies [15]) or strongly bound adatoms [14,20]. To find out the scattering rate associated with these X centers, we introduced an additional concentration of generic RS [16,17] by exposing our devices to atomic hydrogen [14]. The H exposure led to an increase in the D band accompanied by a decrease in $\mu$. The measurements yield that the D peak intensity typically observed in pristine graphene corresponds to the RS concentration sufficient to limit $\mu$ to $\approx 2$ m$^2$/Vs. This finding leaves two alternatives. If a significant part of the X centers in pristine graphene are monovalent defects or vacancies, they must be efficient scatterers similar to H [16,17] and, therefore, constitute the limiting mechanism. However, if the D peak is mostly due to bivalent adatoms (e.g., epoxy groups), which do not result in midgap states [17], the observation of X centers is a red herring, and another mechanism limits $\mu$. In the latter case, the found universal presence of intervalley scatterers is important for understanding of other transport properties of graphene such as, for example, weak localization.

Fig. 1 shows a typical Raman spectrum for pristine graphene obtained by cleavage on an oxidized Si wafer. The two most prominent spectral features are found at ~1580 cm$^{-1}$ and ~2680 cm$^{-1}$ (for green laser wavelength $\lambda$ =514.5 nm) and known as G and G' (or 2D) bands, respectively [18,19,21]. In spectra of defected graphene and at graphene edges, another characteristic feature appears at ~1345 cm$^{-1}$. It requires intervalley scattering and is referred to as the D peak, because it is activated by atomic scale defects. This peak is believed to be negligibly weak or absent in cleaved graphene [18,19].

The inset in Fig. 1 zooms into the D peak region of a typical pristine sample away from the edges. One can see that, however small, the D peak is still present there. To analyze the D peak intensity, we have used the standard quantity $I_D/I_G$ (integrated intensity ratio for D and G bands). For the spectra shown in Fig. 1, $I_D/I_G \approx 1.2\%$ and 32% for pristine and defected graphene, respectively. It is difficult to detect D peaks below a few % in intensity (cf. [15,19]), and we had to integrate for 30 min using a laser power of 1 mW (spot size $\approx 2$ $\mu$m). By varying the power and exposure time, we ruled out any contribution due to laser induced damage. In pristine crystals, we have observed $I_D/I_G$ typically from 0.5% to 1.5% with some samples exhibiting values below 0.5% and up to 3% (we have checked the generality of these observations by measuring samples from two other sources; courtesy of Graphene Industries Ltd. and NTU, Singapore) [22]. Similar D peaks were also found in graphene deposited on surfaces other than SiO$_2$ (PMMA, mica and glass) [10] and in suspended graphene (not annealed by high electric current) [23]. Furthermore, $I_D/I_G$ decreased for bilayer and trilayer graphene (roughly as the number of layers) and was beyond our detection limit ($\approx 0.1\%$) in multilayer samples. This suggests that X centers are located at the surface (that is, they are adsorbates). Annealing at $T$ up to 400°C did not suppress the D peak whereas, at higher $T$, $\mu$ decreased and a broad peak emerged in the D peak region (around ~1360 cm$^{-1}$), possibly because graphene started reacting with the substrate or surroundings.

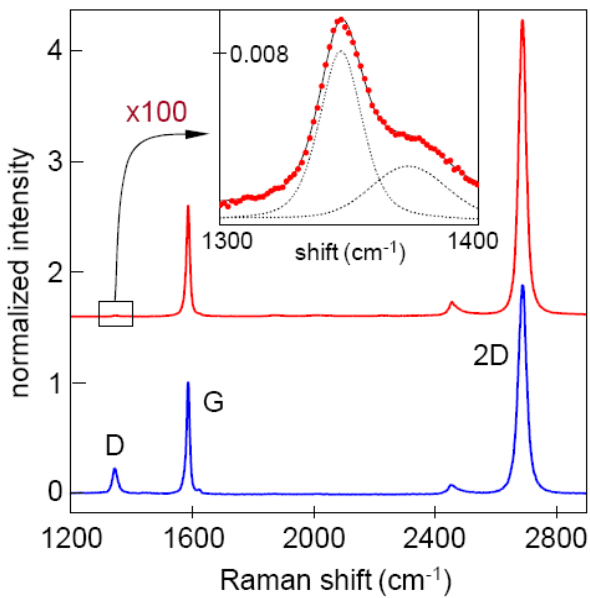

FIG. 1 (color online). Raman spectra for pristine and defected graphene (red and blues curves, respectively; $\lambda$=514.5 nm). In the latter case, the defects were induced by exposure to atomic H. The curves are normalized with respect to the G peak amplitude and shifted for clarity. The inset zooms into the D band region. In our experience, such small D peaks are universally present in Raman spectra of cleaved graphene. In addition to the D peak at ~1345 cm$^{-1}$, an extra feature is present on its right-hand side. The black curve is the best fit of the experimental spectrum (red symbols) by using the Gaussians shown by the dotted curves. The right-side peak does not grow with increasing the number of defects and is clearly resolved for red-laser excitation. Unlike many other small Raman features known for carbon materials [21], the extra peak at 1360 cm$^{-1}$ is not universally present in graphene.

The fact that X centers could not be annealed at 400°C implies their desorption energy >2eV [20], i.e. they are strongly bound adatoms [16,17].

Although the observed D peak unambiguously indicates the presence of intervalley scatterers, it is a priori unclear whether their amount is sufficient to notably affect graphene's transport properties. We can estimate the concentration of X centers as $\approx 10^{10}$ cm$^{-2}$ and $3\times 10^{10}$ cm$^{-2}$ by using the empirical relation found for defects induced by Ar$^+$ bombardment [24] and the Raman spectra presented in another ion bombardment experiment [15], respectively. This yields separation $L$ between X centers $\approx 50$ to 100 nm. The contribution of X centers to electron transport should obviously depend on their scattering efficiency. The latter can also be estimated from Raman data by using [25]

$$L_a(\text{nm}) = (2.4\times 10^{-10})\, \lambda(\text{nm})^4\, (I_D/I_G)^{-1}$$

which yields $L_a \approx 1.7\mu$m for $I_D/I_G =1\%$ and $\lambda =514.5$nm. This widely-used formula was empirically obtained for nanographitic materials, and $L_a$ referred to their crystal sizes. In our case, there are no grain boundaries, and $I_D/I_G$ probes the intervalley rate for randomly distributed defects. Because of the large angle scattering involved, $L_a$ is expected to equal the transport mean free path $l$ for this scattering channel. Typical $l$ for graphene devices are in good agreement with $L$ but an order of magnitude shorter than $L_a$ which at first glance seems to indicate a negligible contribution of X centers into $l$. However, the transport and Raman measurements probe scattering rates at different energies. For the electrons and holes generated at $\lambda =514.5$ nm, their energies ($\approx 1.2$eV) are one order of magnitude larger than typical $E_F$ in transport experiments. In the case of Dirac fermions, one can write $l = \mu E_F/ev_F$ where $v_F$ is the Fermi velocity and $e$ the electron charge. Assuming a constant $\mu$ mechanism and charge carriers with $E_F =1.2$eV and $\mu =2$ m$^2$/Vs, the formula yields $l \approx 2.2\mu$m, in agreement with $L_a$ found above for $I_D/I_G =1\%$. This clearly shows that the intervalley scatterers responsible for the Raman peak become more efficient in transport measurements and may contribute to limited $\mu$.

To find the scattering efficiency of X centers, we mimicked them with H adatoms and studied the relation between the induced $I_D/I_G$ and $\mu$. For these studies we fabricated large Hall crosses such as the one shown in the inset of Fig. 2. The crosses were 4 to 6 μm wide, i.e. significantly larger than the laser spot ($\approx 2\mu$m in diameter) to avoid a Raman signal from sample edges. We employed the bend resistance geometry [26], which allowed us to probe $\mu$ within the same central area that was probed by Raman microscopy. After microfabrication, the crosses exhibited $I_D/I_G >1\%$, even if the original material showed weaker D peaks. This can be attributed to additional X centers induced during microfabrication [27]. For such large devices, mobilities were $\leq 1.2$ m$^2$/Vs with some crosses

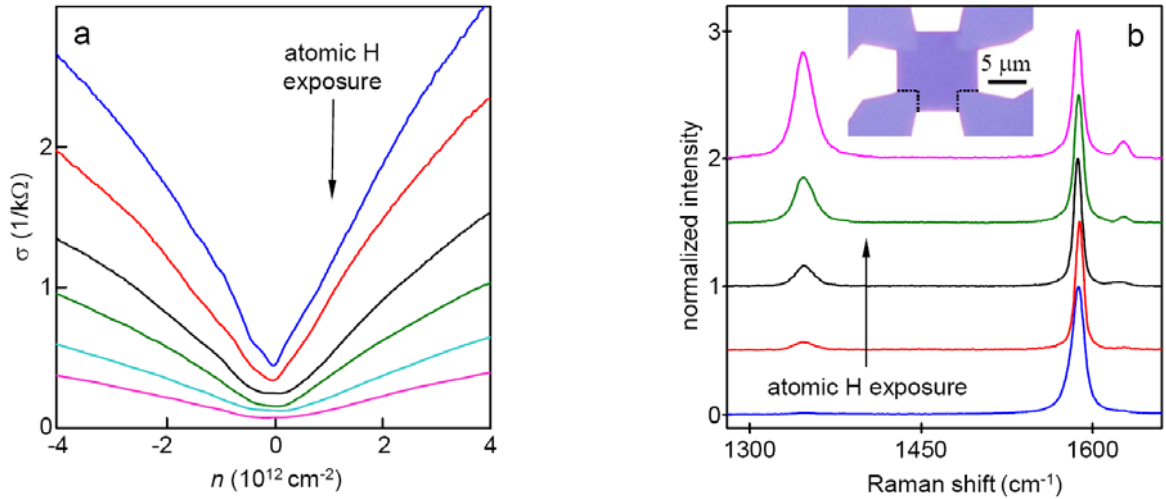

FIG. 2 (color online). Hydrogen adatoms reduce graphene's electronic quality (a) and simultaneously lead to an increase in the D peak (b). Curves of the same color correspond to the same H exposure. The exposure time required to enhance the D peak by the same percentage varied from sample to sample [14]. The transport measurements were typically carried out at 70 K to allow higher electric doping. Carrier concentration $n$ is given with respect to the NP. The gate voltage needed to reach the NP ($n =0$) in (a) varied by less than $\pm 5$ V. The inset shows an optical micrograph of one of the Hall crosses used in these measurements. The thin black lines are to help recognize the device edges.

exhibiting μ down to 0.15 m²/Vs. In the latter devices, we found a significant density of charged impurities, which was verified by immersing the devices in high-κ media [10]. Because of this sample dependence, we did not find any apparent correlation between $I_D/I_G$ and μ in the as-made devices [27].

To proceed further, the Hall crosses were dosed with atomic hydrogen following the procedures described in ref. [14]. H binds to graphene giving rise to a level very close to zero ε (~30 meV) [17]. Fig. 2 shows the evolution of σ(n) and Raman spectra as we increased H exposure. One can see that, as the D peak grows, μ becomes smaller. The observed σ(n) curves were analyzed by separating long- and short- range scattering contributions [7,10]. Figure 3 summarizes our results for 5 different Hall crosses by showing how their long-range $μ_L$ correlated with the induced D peak intensity. One can see that the scattering rate $1/μ_L$ increases approximately linearly with $I_D/I_G$ as expected. Short-range resistivity $ρ_S$ was only weakly affected (see the figure caption).

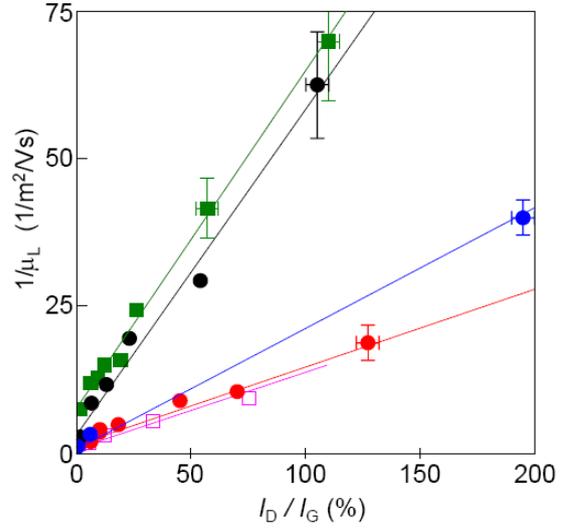

FIG. 3 (color online). Changes in mobility as a function of the D peak intensity. Different symbols denote different devices. Solid lines are the linear fits. The offset along the y-axis in the devices with initially low μ indicates the presence of other types of scatterers (probably, charged impurities) which do not contribute to the D band. We found nearly constant $ρ_S ≈200$ Ω for the devices with the steepest increase in $1/μ_L$. On the contrary, for the two devices with the lowest rate of changes, $ρ_S$ increased from initial ≈100 Ω to 400 Ω at $I_D/I_G$ >100%.

From the slopes of the curves in Fig. 3, we find that, in the absence of other impurities, H adatoms in the concentration that corresponds to $I_D/I_G =1\%$, would limit μ to a value between 1.5 and 8 m²/Vs. The apparently random variations in the observed scattering rate for different devices (by a factor of 5) can be attributed to various degrees of hydrogen clustering. For flat graphene, H tends to make pairs [20,28,29] but this mechanism is suppressed in rippled graphene [29]. H pairs remain intervalley scatterers (i.e. they contribute to the D band and $ρ_S$) but, similar to bivalent adatoms, they do not result in a level near zero ε and are not RS [17,28]. This explanation is supported by the facts that changes in $ρ_S$ a) were noticeable only for the devices with the smallest changes in $μ_L$ and b) tended to be quicker at low $I_D/I_G$. Therefore, the steepest increase in $1/μ_L$ in Fig. 3 would correspond to the case of little clustering and yields such scattering efficiency for isolated monovalent adatoms that $I_D/I_G ≈1\%$ sets up a limit on μ of ≤2 m²/Vs.

The X centers in pristine graphene cannot be H adatoms because the latter are weakly bound and can be removed by annealing at 200°C [14]. One might argue that X centers can have a scattering efficiency very different from H. To this end, we note that the resistivity induced by RS is roughly described by $ρ_D ≈ \frac{h}{e^2} \frac{n_D}{n \ln^2(\sqrt{\pi n} R_D)}$ where $R_D$, $n_D$ and $h/e^2$ are the effective radius of RS, their density and the resistivity quantum, respectively [6]. The logarithmic dependence on $R_D$ assures that different RS contribute rather similarly. Indeed, we have analyzed the Raman spectrum presented in Fig. 1 of ref. [15] for vacancies induced by Ne bombardment and found that their scattering rate corresponds to μ ≈2.5 m²/Vs for $I_D/I_G =1\%$ (converted to λ =514.5 nm), in good agreement with our results for H adatoms. Using the above formula and assuming $R_D ≈2a$ [15] (a =1.42Å is the bond length), we estimate that it requires $n_D <10^{11}$ cm⁻² to limit μ to ≈2 m²/Vs. Therefore, RS are somewhat more efficient scatterers than charged impurities, as it requires the latter in concentrations > $10^{11}$ cm⁻² to set the same limit for μ [4,15,30].

In conclusion, our results provide appealing evidence that the intervalley scatterers responsible for a small D peak in graphene devices can also set up a limit on their μ. The presence of strongly bound monovalent adsorbates or vacancies (in combination with charged impurities that reduce μ even further) can reconcile many previous observations concerning the limiting scattering mechanism.

Unfortunately, it is currently impossible to exclude the alternative that the observed X centers are bivalent or paired adatoms. In this case, they would be non-resonant and contribute relatively little in $\mu$. Nevertheless, they should greatly influence weak localization and $\rho_S$. Further work is required to find out what these X centers are and eliminate them [27]. For the moment, we suggest that they could be due to hydrocarbon contamination (as observed by transmission electron microscopy [31]), which results in occasional C-C bonds to graphene, similar in nature to, for example, $CH_3$ groups [17].